
%
\def\sdp{$$\,\bigcirc\!\!\!\! s\,$$}        
\def\footnoterule{\kern-3pt \hrule width \hsize \kern2.6pt}
\pageno=0                                   
\footline={\ifnum\pageno>0 \hss\folio\hss \else\fi}
\magnification=1200
\baselineskip=20pt
\null
\vskip36pt
\centerline{SPACE-TIME GEOMETRY OF THREE-DIMENSIONAL YANG-MILLS THEORY}
\vskip48pt
\centerline{V. Radovanovi\'c}

\centerline{Faculty of Physics, University of Belgrade, P.O. Box 550,
Belgrade, Yugoslavia}
\vskip12pt
\centerline{and}
\vskip12pt
\centerline{Dj. \v{S}ija\v{c}ki }

\centerline{Institute of Physics, P.O. Box 57, Belgrade, Yugoslavia}

\vskip86pt
\centerline{Abstract:}

{\narrower\smallskip\noindent
It is shown that the $SU(2)$ Yang-Mills theory in $3$-dimensional
Riemann-Cartan space-time can be completely reformulated as a
gravity-like theory in terms of gauge invariant variables. The
resulting Yang-Mills induced equations are found, and it is
demonstrated that they can be derived from a torsion-square type of
action.
\smallskip}

\vfill
\noindent \hrule
\vskip8pt
\noindent E-mail:\quad esijacki@ubbg.etf.bg.ac.yu\quad
radovanovic@castor.phy.bg.ac.yu
\eject
\vskip24pt
{\it 1. Introduction}
Over the past thirty years quite a number of attempts to connect
certain strong interactions phenomena to intricate space-time variables
and/or objects have been suggested.

\noindent (a) Already in 1965, in an attempt to describe the hadron
spectroscopy in terms of non-compact groups, i.e. algebras, Dothan,
Gell-Mann and Ne'eman [1] conjectured that the $\Delta J = 2$ Regge
trajectory rule can be accomplished by making use of the $SL(3,R)$
spectrum-generating algebra (SGA) of operators given as the
time-derivatives of the gravitational quadrupoles. The $SL(3,R)$ SGA
was later combined with the Lorentz algebra yielding the $SL(4,R)$ SGA
[2], thus describing in one shot both the single Regge trajectory
recurrences and the corresponding `daughter' states.

\noindent (b) The Veneziano dual amplitudes of the late sixties were
subsequently rederived in terms of the Nambu-Gotto string. The string
itself, or as an approximation for the gluon-field flux-tubes, opened a
new era of space-time extended objects in Physics [3]. Moreover, the string
indicates that the existence of a gravity-like component within the QCD
gauge should be no surprise -- the truncated massless sector of the
open string reduces to a $J=1$ Yang-Mills field theory, while the same
truncation for the closed string reduces to a $J=2$ gravitational field
theory (since the closed string is the contraction of two open
strings).

\noindent (c) The Bag Model came as another kind of extended object -- it
was proposed as a candidate for the dynamical approximation of confined
color [4]. It is interesting to note that $SL(3,R)$ is the invariance
group of a volume, thus the states described by the corresponding
excitation represent the pulsations and deformational vibrations of a
fixed volume `bag'.

\noindent (d) Salam and others [5-7] have used a gravitational
framework to describe hadrons, assuming it to be genuinely extraneous
to QCD -- perhaps a true short range component of gravity. The `strong
gravity' interactions are mediated through an exchange of the
additional `strong metric', that was originally recognized as the $f_2$
meson (with $J^{P} = 2^{+}$). It has been shown recently that a strong
gravity type of theory -- Chromogravity -- can be constructed, in the
IR approximation, starting from QCD itself [8-9]. In this case one
finds a $p^{-4}$ propagator for the chromometric (two-gluon) field,
indicating confinement, as well as a Regge-like $m^2 \sim J$ spectrum.

\noindent (e) In the course of expressing the QCD theory entirely in
terms of field strengths, Halpern [10] introduced gauge invariant field
variables of a curved space-time. Recently, Lunev [11] considered to
some extent the problem of expressing the $SU(2)$ Yang-Mills theory in
three dimensions in terms of the gauge-invariant variables arriving at
a Rimannian type of geometry. Freedman et al. [12], addressed the
question of the Gauss law constraint in the Hamiltonian form of the
$SU(2)$ gauge theory and found a $3$-dimensional spatial geometry with
$GL(3,R)$ structure that underlines the gauge-invariant configuration
space of the theory.

The aim of this paper is to investigate systematically the possibility
of expressing the $SU(2)$ Yang-Mills theory in $3$-dimensions solely in
terms of $SU(2)$ gauge-invariant space-time field variables, i.e. to
view it as a kind of Riemann-Cartan space-time geometry. In order to
find out the precise role and meaning of the $SU(2)$ induced
gauge-invariant space-time quantities, we find it crucial to embed the
Yang-Mills theory into an extrinsic curved space-time. An efficient way
to do this is to study a gauge theory based on the $[T_{3}\sdp
SO(1,2)]\otimes SU(2)$ group, i.e. to gauge simultaneously both the
$3$-dimensional Poincar\'e symmetry and the $SU(2)$ symmetry.

Let us concentrate first on the $SU(2)$ gauge itself. We will make us
of the gauge potentials $A^{a}_{\mu}$, gauge field strength
$F^{a}_{\mu\nu} = \partial_{\mu} A^{a}_{\nu} - \partial_{\nu}
A^{a}_{\mu} + \epsilon^{abc} A^{b}_{\mu} A^{c}_{\nu}$, i.e.
$F^{a\rho\sigma} = g^{\rho\mu} g^{\sigma\nu} F^{a}_{\mu\nu}$, and its
dual $F^{\star a}_{\mu} = {1\over 2} \epsilon_{\mu\rho\sigma}
F^{a\rho\sigma}$, where $a = 1,2,3$ is the $SU(2)$ index, $\mu ,\nu ,
\dots = 0,1,2$ are the world indices, $\epsilon^{abc}$ are the $SU(2)$
structure constants while $\epsilon_{\mu\rho\sigma}$ ($\epsilon_{012} =
1$) is the Levi-Civita symbol. The inverse $\bar{F}^{\star}$ of the
dual tensor is given by $F^{\star a}_{\mu} \bar{F}^{\star a\nu} =
\delta^{\nu}_{\mu}$. We can now define the following $SU(2)$-scalar
second-rank symmetric tensor:
$$
G_{\mu\nu}
= \eta_{ab} F^{\star a}_{\mu} F^{\star b}_{\nu} det(\bar{F}^{\star}),
\eqno (1.1)
$$
where $\eta_{ab}$ is the $SU(2)$ Cartan metric tensor, and
$det(\bar{F}^{\star}) = det(\bar{F}^{\star a\mu})$. The starting
Yang-Mills theory (expressed, say, in terms of $A^{a}_{\mu}$) has $6$
physical degrees of freedom that equals the number of $G_{\mu\nu}$
components. This fact as well as the fact that the $SU(2)$ group can be
naturally mapped onto the continuous part of the automorphisms group of
the $3$ space-time translations are in the root of the geometric
reformulation of the $SU(2)$ Yang-Mills theory in terms of gauge
invariant variables. In the space-time reformulation, the $G_{\mu\nu}$
tensor plays the role of a metric tensor of the `Yang-Mills induced'
Riemman-Cartan-like geometry.

\vskip24pt
{\it 2. Gauging and the transformation laws.}
Let us consider a gauge theory based on the $[T_{3}\sdp SO(1,2)]\otimes
SU(2)$ group. The local (anholonomic) $3$-dimensional Lorentz indices
are denoted by $i, j, k, \dots$, while the world (holonomic) indices
are denoted by the Greek letters. Moreover, we denote the genuine
gravity variables by small letters, whereas the Yang-Mills and/or new
gravity-like Yang-Mills induced variables we denote by capital letters.
The covariant derivatives generating a parallel transport of a
tangent-space matter-field $\psi$ are given as follows:
$$
\matrix{
SU(2)\ :\hfill & D_{i} \psi = e^{\mu}_{i} (\partial_{\mu}
+ iA^{a}_{\mu}T^{a}) \psi \hfill \cr
SO(1,2)\ :\hfill & \tilde{D}_{i} \psi = e^{\mu}_{i} (\partial_{\mu}
+ {i\over 2}a^{jk}_{\mu}S_{jk}) \psi \hfill \cr
SO(1,2)\otimes SU(2)\ : & \bar{D}_{i} \psi = e^{\mu}_{i} (\partial_{\mu}
+ {i\over 2}a^{jk}_{\mu}S_{jk} + iA^{a}_{\mu}T^{a}) \psi , \cr}
\eqno (2.1)
$$
where $S_{jk}\ (i,j = 1,2,3)$, and $T^{a}\ (a = 1,2,3)$ are the
Lorentz, and the $SU(2)$ group generators respectively. Moreover,
$e^{i}_{\mu}$ (the triads), $a^{ij}_{\mu}$, and $A^{a}_{\mu}$ are the
translational, Lorentz, and $SU(2)$ gauge potentials respectively. The
gauge potentials (infinitesimal) transformation law is given by
$$
\eqalign{
\delta e_{i}^{\mu}(x) &= \omega_{i}^{\ j} e_{j}^{\mu}(x)
+ \xi^{\mu}_{,\nu} e_{i}^{\nu} \cr
\delta a^{ij}_{\mu}(x) &= \omega^{i}_{\ k} a^{kj}_{\mu}
+ \omega^{j}_{\ k} a^{ik}_{\mu}
- \xi^{\nu}_{,\mu} a^{ij}_{\nu} - a^{ij}_{,\mu} \cr
\delta A^{a}_{\mu}(x) &=
\partial_\mu \theta^{a} + \epsilon^{abc} \theta^{b} A^{c}_{\mu}
- \xi^{\nu}_{,\mu} A^{a}_{\nu}, \cr}
\eqno (2.2)
$$
where $\xi^{\mu}$, $\omega^{ij}$, and $\theta^{a}$ are the group
parameters corresponding to the translational $T_{3}$, Lotentz
$SO(1,2)$, and $SU(2)$ groups respectively. The translational gauge
invariance requires the action integration measure $d^{3}x$ to be
replaced by $d^{3}x\sqrt{g}$, where $g = det(g_{\mu\nu})$, and where
$g_{\mu\nu} = \eta_{ij} e^{i}_{\mu} e^{j}_{\nu}$ ($\eta_{ij} =
diag(+1,-1,-1)$) is the genuine space-time metric.

The Levi-Civita symbol $\epsilon_{\mu\nu\rho}$ is a tensor-density with
respect to the general coordinate transformations $x \rightarrow
x'(x)$, i.e.
$$
\epsilon{}'_{\mu\nu\rho} =
\left\vert {\partial x'\over \partial x}\right\vert
{\partial x^\lambda\over \partial x^{'\mu }}
{\partial x^\eta\over \partial x'^\nu }
{\partial x^\kappa\over \partial x'^\rho }
\epsilon_{\lambda\eta\kappa }.
\eqno (2.3)
$$
Therefore, the $SU(2)$ field strength dual tensor $F^{\star a}_{\mu}$
transforms as
$$
F^{'\star a}_{\mu} =
\left\vert {\partial x'\over \partial x}
\right\vert {\partial x^\nu\over \partial x'^\mu}
F^{\star a}_{\nu},
\eqno (2.4a)
$$
implying
$$
det(\bar{F}^{\star}) \rightarrow
\left\vert {\partial x\over \partial x'}\right\vert^2
det(\bar{F}^{\star}),
\eqno (2.4b)
$$
and thus we can define the following world vectors
$$
E^{a}_{\mu} = F^{\star a}_{\mu} \sqrt{det(\bar{F}^{\star})},
\eqno (2.5)
$$
that play the role of {\it Yang-Mills induced triads}, i.e. they convert
mutually the Yang-Mills and world indices. Now we can rewrite (1.1) as
follows
$$
G_{\mu\nu} = \eta_{ab} E^{a}_{\mu} E^{b}_{\nu}.
\eqno (2.6)
$$
The corresponding $G_{\mu\nu}$ general-coordinate transformation-rule is
$$
G'_{\mu\nu}(x') = {\partial x^\rho \over \partial x'^\mu }
{\partial x^\sigma \over \partial x'^\nu} G_{\rho\sigma}(x),
\eqno (2.7)
$$
and thus the $G_{\mu\nu}(x)$ tensor field represent the {\it
gauge-invariant Yang-Mills induced space-time metric}.

Following [11, 12], we define the {\it gauge-invariant Yang-Mills
induced space-time connection} as follows, $$ D_{\mu} E^{a}_{\nu} =
\Gamma^{\rho}_{\mu\nu} E^{a}_{\rho},
\eqno (2.8)
$$
where $D_{\mu} = e^{i}_{\mu} D_{i} = \partial_{\mu} + [A_\mu ,\ \  ]$
is the $SU(2)$ covariant derivative [cf (2.1)]. Rewriting (2.8) as
$$
\partial_{\mu} E^{a}_{\nu} + \epsilon^{abc} A^{b}_{\mu} E^{c}_{\nu}
- \Gamma^{\rho}_{\mu\nu} E^{a}_{\rho} = 0,
\eqno (2.9)
$$
we find that $A^{ab}_{\mu} \equiv - \epsilon^{abc} A^{c}_{\mu}$ plays
exactly the role that has the spin connection in a Poincar\'e gauge
theory.  Equation (2.9) relates the spin-like and affine-like
Yang-Mills induced connections. To sum up, the Yang-Mills potentials
$A^{ab}_{\mu}$ and the properly normalized dual field strengths
$E^{a}_{\mu}$ appear geometrically in the same way as the connections
and triads of the configuration space-time appear respectively. By
making use of (2.8) and (2.2), we find the affine connection
transformation properties with respect to the general coordinate
transformations as:
$$
\Gamma^{'\mu }_{\nu\rho} =
{\partial x^{'\mu }\over \partial x^{\alpha }}
{\partial x^\beta \over \partial x^{'\nu }}
{\partial x^\gamma \over \partial x^{'\rho }}
\Gamma^\alpha_{\beta\gamma}
+ {\partial^{2} x^\lambda\over \partial x^{'\nu} \partial x^{'\rho }}
{\partial x^{'\mu }\over \partial x ^\lambda }.
\eqno (2.10)
$$

Let us restrict now to the $SU(2)$ gauge itself. From (2.2) one has
$\delta F^{\star a}_{\mu} = \epsilon^{abc}\theta^{b}F^{\star
c}_{\mu}$ and $D'_\mu E^{'a}_{\nu} = D_\mu E^a_\nu +
\epsilon^{abc}\partial_\mu E^b_\nu \theta^c + o(\theta^2)$ as well as
$\delta_{YM} det(\bar{F}^{\star}) = 0$, and thus we find, as expected,
that
$$
\eqalign{
\delta_{YM}& G_{\mu\nu} = 0, \cr
\delta_{YM}& \Gamma^\mu_{\nu\rho} = 0. \cr}
\eqno (2.11)
$$
Multiplying (2.8) with $E^{a}_{\sigma}$ and symmetrizing ($\sigma\
\leftrightarrow\  \nu$) we have
$$
\nabla_\mu G_{\sigma\nu} \equiv \partial_\mu G_{\sigma\nu}
- \Gamma^\rho_{\mu\nu} G_{\rho\sigma}
- \Gamma^\rho_{\mu\sigma} G_{\rho\nu} = 0.
\eqno (2.12)
$$
The meaning of this relation is the compatibility of the Yang-Mills
induced connection with the corresponding metric. In other words {\it
the space-time geometry defined by $G_{\mu\nu}$ and
$\Gamma^\rho_{\mu\nu}$ is of the Riemann-Cartan type}.

\vskip24pt
{\it 3. Curvature and Torsion}
In building up the space-time Yang-Mills induced geometrical structure
we can introduce respectively the corresponding curvature and torsion
tensors:
$$
\eqalign{
&R^\rho_{\sigma\mu\nu} =
\partial_\mu \Gamma^\rho_{\nu\sigma }
- \partial_\nu \Gamma^\rho_{\mu\sigma }
+ \Gamma^\rho_{\mu\alpha}\Gamma^\alpha_{\nu\sigma}
- \Gamma^\rho_{\nu\alpha}\Gamma^\alpha_{\mu\sigma }, \cr
&T^\mu_{\nu\rho} =
\Gamma^\mu_{\nu\rho} - \Gamma^\mu_{\rho\nu}
= E^{a\mu} (\partial_\nu E^a_\rho - \partial_\rho E^a_\nu
+ A^{ac}_\nu E^c_\rho - A^{ac}_\rho E^c_\nu ), \cr}
\eqno (3.1)
$$
where $E^{a\mu} = G^{\mu\nu} E^a_{\nu}$. We raise and lower indices of
the Yang-Mills originated quantities by the corresponding metric
$G_{\mu\nu}$. The Riemann tensor evaluation is defined by (1.1) and
(2.8) and this tensor can be determined completely at the kinematical
level, while the torsion tensor, as will be seen below, is related to
the Yang-Mills equations of motion.

We can find $R^\rho_{\sigma\mu\nu}$ by evaluating $[D_\mu ,D_\nu ]$ in
two ways. On one hand, using $D_\mu E^{a\nu} = - \Gamma^\nu_{\mu\rho }
E^{a\rho}$ [cf. (2.8)], we have
$$
\eqalign{
[D_\mu ,D_\nu ] E^{a\rho}
&= \Gamma^\rho_{\nu\alpha} D_\mu E^{a\alpha}
+ \Gamma^\rho_{\mu\alpha} D_\nu E^{a\alpha}
- E^{a\alpha}\partial_\mu \Gamma^\rho_{\nu\alpha}
+ E^{a\alpha}\partial_\nu \Gamma^\rho_{\mu\alpha} \cr
&= E^{a\tau} (\partial_\mu \Gamma^\rho_{\nu\tau}
- \partial_\nu \Gamma^\rho_{\mu\tau}
+ \Gamma^\rho_{\mu\alpha} \Gamma^\alpha_{\nu\tau}
- \Gamma^\rho_{\nu\alpha} \Gamma^\alpha_{\mu\tau})
= R^{\rho}_{\tau\mu\nu} E^{a\tau}, \cr }
\eqno (3.2)
$$
while on the other hand we find
$$
\eqalign {
[D_\mu ,D_\nu ] E^{a\rho}
&= D_\mu (\partial _\nu E^{a\rho} + \epsilon^{abc} A^b_\nu E^{c\rho})
- D_\nu (\partial _\mu E^{a\rho} + \epsilon^{abc} A^b_\mu E^{c\rho})\cr
&= -\epsilon^{abc} F_{\mu\nu}^b E^{c\rho}. \cr }
\eqno (3.3)
$$
Equating this two expressions we arrive at $R^\rho_{\tau\mu\nu}
E^{a\tau} = - \epsilon^{abc} F_{\mu\nu}^b E^{c\rho}$. Multiplying this
relation by $E^{a}_{\sigma}$, we have $R^\rho_{\sigma\mu\nu} = -
\epsilon^{abc} F^b_{\mu\nu} E^{c\rho} E^a_\sigma$ $=$ $- \epsilon^{abc}
\bar{F}^{\star c\rho} F^{\star a}_{\sigma} g_{\mu\alpha} g_{\nu\beta}
F^{\alpha\beta b}$ $=$ \hfill\break $- \epsilon^{abc}
\bar{F}^{\star c\rho} F^{\star a}_{\sigma} g_{\mu\alpha} g_{\nu\beta}
\epsilon^{\alpha\beta\kappa} F^{\star b}_{\kappa}$ $=$ $-{1\over {G}}
\epsilon_{\sigma\kappa\tau} \epsilon^{\alpha\beta\kappa}
\bar{F}^{\star c\tau} \bar{F}^{\star c\rho} g_{\mu\alpha} g_{\nu\beta}$,
where $G = det(G_{\mu\nu})$, and finally we find the Yang-Mills induced
Riemann tensor as follows
$$
R^\rho_{\sigma\mu\nu}
= G^{\rho\tau} (g_{\mu\sigma} g_{\nu\tau} - g_{\mu \tau }g_{\nu\sigma}).
\eqno (3.4)
$$
The Yang-Mills induced Ricci tensor reads
$$
R_{\sigma\nu} = \delta^\mu_\rho R^\rho_{\sigma\mu\nu} =
G^{\rho\tau} (g_{\rho\sigma} g_{\nu\tau} - g_{\rho\tau} g_{\nu\sigma}),
\eqno (3.5)
$$
while the Yang-Mills induced scalar curvature takes the following form
$$
R = G^{\sigma\nu}R_{\sigma\nu} = - 2{g\over {G}}g^{\mu\sigma}G_{\mu\sigma}.
\eqno (3.6)
$$

The gauge algebraic structure of relevant quantities is illustrated by
the following diagram:
$$
\matrix{
& & T_{3} & SO(1,2) & SU(2) & & & & \cr
& & \downarrow & \downarrow & \downarrow & & & & \cr
g_{\mu\nu} & \Leftarrow & e^{i}_{\mu} & \qquad a^{ij}_{\mu}\
(\gamma^{\rho}_{\mu\nu}) \qquad &
A^{a}_{\mu}\ (\Gamma^{\rho}_{\mu\nu}) & \rightarrow & E^{a}_{\mu} &
\Rightarrow & G_{\mu\nu} \cr
& & \vert & \vert & \vert & & \vert & & \cr
& & \vert  & \tilde{D}_{\mu}\ (\tilde{\nabla}_{\mu}) & D_{\mu}\
(\nabla_{\mu}) & &\vert & & \cr
& & \downarrow & \downarrow & \downarrow & & \downarrow & & \cr
& & t^{\rho}_{\mu\nu} &   r^{\rho}_{\sigma\mu\nu} & R^{\rho}_{\sigma\mu\nu}
& &  T^{\rho}_{\mu\nu} & & \cr}
$$
where, $e^{i}_{\mu}$, $g_{\mu\nu}$, $a^{ij}_{\mu}$,
$\gamma^{\rho}_{\mu\nu}$, $t^{\rho}_{\mu\nu}$, and
$r^{\rho}_{\sigma\mu\nu}$ refer to the true-gravity triads, metric,
spin and affine connection, torsion and curvature respectively.

\vskip24pt
{\it 4. Space-time form of the Yang-Mills field equations}
The Yang-Mills action in the external Riemann-Cartan space-time has the
following well known form
$$
S_{YM} = - {1\over {4}} \int dx \sqrt{g}
F^a_{\mu\nu} F^a_{\rho\sigma} g^{\mu\rho} g^{\nu\sigma}.
\eqno (4.1)
$$
Variation of (4.1) with respect to $A^a_\mu$ yields the curved-space
Yang-Mills field equations $\partial_\sigma (\sqrt{g} F^{a\sigma\rho})
+ \sqrt{g} \epsilon^{abc} A^b_\sigma F^{c\sigma\rho} = 0$, that can be put
in the following form
$$
D_{\sigma} F^{a\sigma\rho} + \left\{ {\sigma\atop {\sigma\tau}}\right\}_g
F^{a\tau\rho} = 0.
\eqno (4.2)
$$
These equations can be expressed solely in terms of the new
space-time quantities. Substituting $F^{a\rho\sigma} =
\epsilon^{\rho\sigma\mu} F^{\star a}_{\mu}$ in (4.2), making use of
the fact that (true gravity) Christoffel symbol is symmetric in lower
indices, antisymmetraizing in $(\sigma,\ \mu )$, and multiplying by
$\epsilon_{\rho\alpha\beta}$ we find
$$
D_{\sigma} F^{\star d}_{\mu} - D_{\mu} F^{\star d}_{\sigma}
+ \left\{{\rho\atop {\sigma\rho}}\right\}_g F^{\star d}_{\mu}
- \left\{{\rho\atop {\mu\rho}}\right\}_g F^{\star d}_{\sigma} = 0.
\eqno  (4.3)
$$
 From (2.8) one has $D_\mu F^{\star a}_{\nu} = \Gamma^{\rho}_{\mu\nu}
F^{\star a}_{\rho} - {1\over 2} F^{\star a}_{\nu}
{\partial_\mu G\over G}$. Substituting this expression in (4.3),
multiplying by $\bar{F}^{\star d\alpha}$, and by expressing the torsion
tensor in terms of the affine connections (cf. (3.1)), we arrive at
$$
T^\rho_{\sigma\mu}
- {1\over 2}{g\over G}\partial_\sigma {G\over g} \delta^\rho_\mu
+ {1\over 2}{g\over G}\partial_\mu {G\over g} \delta^\rho_\sigma =0 .
\eqno (4.4)
$$
The $\rho ,\mu$ indices contraction in (4.4) implies the following
expression for the contracted torsion
$$
T_{\sigma} \equiv T^\rho_{\sigma\rho} = \partial_\sigma ln{G\over g}.
\eqno (4.5)
$$
Finally, from (4.4) and (4.5) we obtain
$$
T^\rho_{\sigma\mu}
- {1\over 2}\delta^\rho_\mu T_\sigma
+ {1\over 2}\delta^\rho_\sigma T_\mu =0.
\eqno (4.6)
$$
Thus, {\it the Yang-Mills field equations can be   expressed solely in
terms of the new gauge-invariant space-time quantities as given by
(4.6) and the condition (4.5)}. We point out that eq. (4.6) is of the
kind found in a $R + T^2$ theory of gravity. Indeed, we will
demonstrate below that (4.6) can be derived from an appropriate action
expressed in terms of new space-time variables only. At first glance it
seems (in the absence of the matter field source on the right hand
side) that eq. (4.6) yields vanishing torsion. However, its vector
component is given by (4.5), while the other two irreducible components
vanish. Thus, {\it the Yang-Mills induced torsion is non-vanishing}.

A trial Yang-Mills charged (or composite, `color' neutral) particle
experiences two metrics, the true gravity one $g_{\mu\nu}$ and the
Yang-Mills induced one $G_{\mu\nu}$, as seen from (4.2) rewritten as
follows
$$
\partial_\sigma F^{\mu\sigma\rho}
+ \left\{{\sigma\atop {\sigma\nu}}\right\}_g F^{\mu\nu\rho}
+ \left\{{\rho\atop {\sigma\nu}}\right\}_g F^{\mu\sigma\nu}
+ \Gamma^\mu_{\sigma\nu} F^{\nu\sigma\rho} = 0,
\eqno (4.7)
$$
where $F^{\mu\sigma\rho} \equiv F^{d\sigma\rho} E^{d\mu}$. The
$SU(2)$-singlet gauge and `lepton-like' particles experience
$g_{\mu\nu}$ only.

\vskip24pt
{\it 5. Bianchi identities.}
The commutator of two most general covariant derivatives $\bar{D}_{i}$
of (2.1) reads
$$
[\bar{D}_{i}, \bar{D}_{j}] =
t^{\mu}_{ij} \bar{D}_{\mu} + r^{mn}_{ij} S_{mn} + F^{a}_{ij} T^{a}.
\eqno (5.1)
$$
The gauge algebra closure is achieved by requiring the Jacobi identity,
i.e. $[\bar{D}_{i}, [\bar{D}_{j}, \bar{D}_{k}]] + cycl (i,j,k) = 0$,
that yields the following Bianchi identities
$$
\eqalign{
&\tilde{D}_{\mu} t^i_{\nu\rho} + cycl(\mu ,\nu ,\rho )
= r^i_{\mu\nu\rho}, \cr
&\tilde{D}_{\mu} r^j_{\nu\rho i} + cycl(\mu ,\nu ,\rho ) =0, \cr
&D_\mu F^a_{\nu\rho} + cycl(\mu ,\nu ,\rho ) = 0. \cr}
\eqno (5.2)
$$
Let us rewrite the third set of equations of (5.2) in terms of the
Yang-Mills induced variables. By making use of $D_{\mu} F^{a}_{\nu\rho}
= D_{\mu} (\epsilon_{\nu\rho\sigma} g g^{\gamma\sigma}
F^{\star a}_{\gamma})$, the Yang-Mills Bianchi becomes
$$
D_{\mu} F^{a}_{\nu\rho} + cycl(\mu ,\nu ,\rho ) =
D_{\mu} (g\epsilon_{\nu\rho\sigma} g^{\sigma\gamma} F^{\star a}_{\gamma})
+ cycl(\mu ,\nu ,\rho ).
\eqno (5.3)
$$
Multiplying (3.9) by $\epsilon^{\mu\nu\rho}$ we get
$D_{\mu} (g g^{\mu\nu} F^{\star a}_{\nu}) = 0$, implying
$$
(\partial_\mu g) g^{\mu\nu} F^{\star a}_\nu + g(\partial_\mu g^{\mu\nu})
F^{\star a}_\nu
+ (D_\mu F^{\star a}_\nu )g g^{\mu\nu}=0.
\eqno (5.4)
$$
It is easy to check (due to contorsion antisymmetry) that
$\partial_{\mu} ln(\sqrt{G}) = \Gamma^{\rho}_{\mu\rho}$
and
$\partial_{\mu} ln(\sqrt{g}) = \gamma^\rho_{\mu\rho}$, giving
$$
\partial_{\mu} g^{\mu\nu} + \Gamma^{\nu}_{\mu\rho} g^{\mu\rho} -
\Gamma^{\rho}_{\mu\rho} g^{\nu\mu} + 2\gamma^{\rho}_{\mu\rho} g^{\nu\mu}=0.
\eqno (5.5)
$$
 From (4.5) we get $\Gamma^{\rho}_{\sigma\rho} +
\Gamma^{\rho}_{\rho\sigma} = 2 \gamma^{\rho}_{\sigma\rho}$, and thus we
finally obtain the Yang-Mills Bianchi identities in terms of the new
gauge-invariant variables in the following form
$$
\nabla_{\mu} g^{\mu\nu} \equiv \partial_{\mu} g^{\mu\nu}
+ \Gamma^{\nu}_{\mu\sigma} g^{\mu\sigma}
+ \Gamma^{\mu}_{\mu\sigma} g^{\nu\sigma} =0.
\eqno (5.6)
$$
This expression states that {\it the contracted metricity of the
genuine space-time metric $g_{\mu\nu}$ with respect to the Yang-Mills
induced connection $\Gamma^{\rho}_{\mu\nu}$ vanishes}. Note that an
analogous relation is required in the bimetric theory of gravity as an
additional constraint relating two kinds of gravity [13]. In our
case, (5.6) is a kinematical constraint.

\vskip24pt
{\it 6. The Action.}
The Yang-Mills field equations, rewritten in terms of new variables, as
given by (4.6) mimic the form of the $R + T^2$ Poincar\'e gauge theory
equations. We will show now that these equations can be obtained from
the following action ($e$ is the interaction constant)
$$
\eqalign{
S =
&{1\over e^2} \int dx \sqrt{g} (-{1\over 4}
T_{\mu\nu\rho}T^{\mu\nu\rho}
+ {1\over 2} T_{\mu\nu\rho}T^{\nu\rho\mu}
+ {1\over 2}T_\nu T^\nu ) \cr
&+ \int dx \sqrt{g} \lambda^{\mu\tau}
(E^{a}_{\mu}-{1\over 2} \epsilon_{\mu\nu\rho}
F^{a\nu\rho} \sqrt{G})
(E^{a}_{\tau}-{1\over 2}\epsilon_{\tau\sigma\kappa}
F^{a\sigma\kappa} \sqrt{G}).
 \cr }
\eqno (6.1)
$$
The last term in (6.1) is of no importance in derivation of the new
form of the Yang-Mills equations (4.6). However this term allows us to
transform (4.6) back to the original Yang-Mills form.

The variation of (6.1) with respect to $A^{ab}_{\mu}$ gives
$$
\eqalign{
- E^{a\nu} E^{b\rho} {\delta S\over \delta A^{ab}_\mu }
=& T^{\mu\nu\rho} + {1\over 2} (G^{\mu\nu} T^\rho
- G^{\mu\rho} T^\nu) \cr
&+ \lambda^{\alpha\tau} (E^{c}_{\alpha} -
{1\over 2} \epsilon_{\alpha\beta\gamma} F^{c\beta\gamma} \sqrt{G})
( - {1\over 2} \sqrt{G} \epsilon_{\tau\sigma\kappa}
{\delta F^{c\sigma\kappa}\over \delta A^{ab}_{\mu}})
E^{a\nu} E^{b\rho} = 0. \cr}
\eqno (6.2)
$$

The variation of (6.1) with respect to $E^{a}_{\alpha}$ yields
$$
\eqalign{
E^a_\lambda {\delta S\over \delta E^a_\alpha }
&= {\sqrt{g}\over e^2} \{-D_\epsilon (T^{\mu\epsilon\alpha} E^a_\mu )
E^a_\lambda
+ T^{\alpha\nu\rho} T_{\lambda\nu\rho}
+ T_{\mu\lambda\rho} T^{\mu\alpha\rho} \cr
&+ D_\nu (T^{\nu\alpha\mu} E^a_\mu ) E^a_\lambda
- D_\rho (T^{\alpha\rho\mu} E^a_\mu) E^a_\lambda
- T^{\alpha\rho\mu} T_{\mu\lambda\rho} \cr
& -2T^{\rho\alpha\nu} T_{\lambda\nu\rho}
- T^{\mu\nu\alpha} T_{\nu\lambda\mu}
+ D_\epsilon (T^\epsilon E^{a\alpha}) b^a_\lambda \cr
&- D_\epsilon (T^\alpha E^{a\epsilon}) E^a_\lambda
- 2T^\beta T_{\lambda\beta}^{\ \ \alpha}
- T^{\alpha\nu}_{\ \ \lambda} T_\nu - T^\alpha T_\lambda \} \cr
&+ {\partial_\epsilon \sqrt{g}\over e^2}
(T^{\epsilon\alpha\mu} G_{\mu\lambda}
- T^{\alpha\epsilon\mu} G_{\mu\lambda}
-T^{\mu\epsilon\alpha} G_{\mu\lambda}
+ T^\epsilon \delta^\alpha_\lambda
- T^\alpha \delta^\epsilon_\lambda ) \cr
&-2E^{a}_{\lambda} \sqrt{g}
(E^{b}_{\mu} - {1\over 2} \epsilon_{\mu\nu\rho} F^{b\nu\rho} \sqrt{G})
(\delta^\alpha_\tau \delta^{ab} - {1\over 2} \sqrt{G} E^{a\alpha}
\epsilon_{\tau\sigma\kappa} F^{b\sigma\kappa}) \lambda^{\mu\tau} = 0 \cr}
\eqno (6.3)
$$

And finally, variation of (6.1) with respect to the Lagrange multipliers
$\lambda^{\mu\tau}$ implies
$$
{\delta S\over \delta\lambda^{\mu\tau}} =
\sqrt{g}(E^{a}_{\mu} -
{1\over 2}\epsilon_{\mu\nu\rho}F^{a\nu\rho} \sqrt{G})
(E^{a}_{\tau} -
{1\over 2}\epsilon_{\tau\sigma\kappa}F^{a\sigma\kappa} \sqrt{G}) = 0.
\eqno (6.4)
$$
Equation (6.4) can be factorized (with an appropriate orthogonal
transformation) producing the following relation
$$
E^{a}_{\mu} - {1\over 2} \epsilon_{\mu\nu\rho} F^{a\nu\rho} \sqrt{G} = 0.
\eqno (6.5)
$$

First, we substitute (6.5) into (6.2) and (6.3) getting rid of the terms
proportional to $\lambda$. Second, we substitute thus modified (6.2)
into (6.3) arriving at an identity ($0 = 0$), and finally we are left
with the modified (no $\lambda$ terms) equation (6.2) that is identical
to (4.4).

The main results of this paper can be schematically presented as follows:
$$
- {1\over 4} \int dx \sqrt{g} F^2 \quad
\Longrightarrow \quad (4.2) \quad
\longleftrightarrow \quad (4.6) \quad
\Longleftarrow \quad
- {1\over e^2} \int dx \sqrt{g} T^2 .
$$
In other words, on one hand we start from a $3$-dimensional $SU(2)$
gauge theory in an external Riemann-Cartan space-time (or a flat
space-time and gauged Poincar\'e symmetry), and rewrite the theory in
terms of gauge invariant variables (the Yang-Mills induced metric and
connection). The original Yang-Mills equations (4.2) are shown to be
given in terms of the induced torsion, eq. (4.6), while the
Yang-Mills Bianchi identities are given as the Poincar\'e metricity
condition for the gauge invariant connection (5.6). On the other hand
we start with a gravity-like torsion-square action (6.1), and derive
eq. (4.6), that can be (making use of the Lagrange multipliers given
relations) recast back into the original Yang-Mills form.

\vfill
\eject

{\it References:}

\item{1.}
{Y. Dothan, M. Gell-Mann and Y. Ne'eman, {\it Phys. Lett.} {\bf 17}
(1965) 148.}

\item{2.}
{Dj. \v{S}ija\v{c}ki, Ph.D Thesis, Duke University 1974).}

\item{3.}
{M. Jacob, {\it Dual Theory}, (North-Holland, 1974); J. Scherk, {\it
Rev. Mod. Phys.} {\bf 47} (1975) 123; M.B. Green, J.H. Schwarz and E.
Witten, {\it Superstring Theory}, (Cambridge U.P., 1987).}

\item{4.}
{A. Chodos, et al. {\it Phys. Rev. D} {\bf 9} (1974) 3471.}

\item{5.}
{C.J. Isham, A. Salam and J. Strathdee, {\it Phys.  Rev. D} {\bf 8}
(1973) 2600; {\bf 9} (1974) 1702; C. Sivaram and K.  Sinha, {\it Phys.
Rep.} {\bf 51} (1979) 111.}

\item{6.}
{Y. Ne'eman and Dj. \v Sija\v cki, {\it Ann. Phys. (NY)} {\bf 120}
(1979) 292.}

\item{7.}
{P. Caldirola, M. Pav\v si\v c and E. Recami, {\it Nuovo Cimento B}
{\bf 48} (1978) 205; {\it Phys. Lett.} {\bf 66A} (1978) 9.}

\item{8.}
{Dj. \v{S}ija\v{c}ki, Y. Ne'eman, {\it Phys. Lett. B} {\bf 247} (1990)
571.}

\item{9.}
{Y. Ne'eman, Dj. \v{S}ija\v{c}ki, {\it Phys. Lett. B} {\bf 276B} (1992)
173.}

\item{10.}
{M. B. Halpern, {\it Phys. Rev. D,} {\bf 16} (1977) 1798; {\bf 16}
(1977) 3515; {\bf 19} (1979) 517.}

\item{11.}
{F.A. Lunev, {\it Phys. Lett.} {\bf 295B} (1992) 92.}

\item{12.}
{D.Z. Freedman, P.E. Haagensen, K. Johnson and J.I. Latorre, MIT CTP-2238
preprint (August 1993); M. Bauer, D.Z. Freedman and P.E. Haagensen,
CERN-TH.7238/94 preprint (May 1994).}

\item{13.}
{N. Rosen, in {\it Cosmology and Gravitation}, P.G. Bergman and V.de
Sabbata eds., (Plenum Press, 1980).}

\end